\documentclass[preprint,aps,prc,superscriptaddress,showpacs,showkeys]{revtex4}  
\usepackage{times}
\usepackage[dvips]{graphicx}
\newcommand{\be}{\begin{eqnarray}}
\newcommand{\ee}{\end{eqnarray}}

\begin{document}
\title{Comment on the paper "Energy Loss of Charm Quarks in the Quark-Gluon Plasma :
Collisional vs Radiative"}

\vskip 0.2in

\author{M. Mishra}
\email{Email: madhukar.12@gmail.com}
\affiliation{Department of Physics, Banaras Hindu University, Varanasi 221005, India}
\author{V. J. Menon}
\affiliation{Department of Physics, Banaras Hindu University, Varanasi 221005, India}
\author{B. K. Patra} 
\affiliation{Indian Institute of Technology Roorkee, Roorkee 247667, India}
\author{R. K. Dubey}
\affiliation{Department of Physics, Banaras Hindu University, Varanasi 221005, India}

\vskip 0.2in

\begin{abstract}
  In the article by M. G. Mustafa published in Phys. Rev. C {\bf 72}, 014905 (2005) the author has estimated the total energy loss of a charm quark and quenching of hadron spectra due to the collisional energy loss of energetic partons in an expanding quark-gluon plasma employing Fokker-Planck equation. We wish to point out through this Comment that some of conceptual and numerical results of the said paper are unreliable.
For the sake of clarity our discussion will focus on the massless case (although a few remarks on the $m\neq 0$ case are also made).
\end{abstract} 
\vskip 0.9in



\maketitle

\section*{General remarks}
The said paper by Mustafa~\cite{must} and an earlier article by Mustafa and Thoma~\cite{mustho} were aimed at expanding further the pioneering works by Baier et al.,~\cite{bair} and M\"{u}ller~\cite{mull} on the important phenomenon of jet quenching, i.e., the suppression of hadronic spectra produced by partons (having large transverse momentum) which suffer sizable energy loss while traveling in a hot and dense quark-gluon plasma. Refs.~\cite{must,mustho} attempted to fulfil this aim by explaining the relevant physical mechanisms, recalling the expressions of collisional \& radiative energy loss rates predicted by finite-temperature QCD, setting up a Fokker-Planck transport equation for massless or massive partons, solving the partial differential equation analytically by using Fourier transform together with the method of characteristics, and plotting the predicted quenching factor $Q(p_T)$ and $D/\pi$ ratio graphically for a Bjorken expanding QGP expected to have been produced in ultra-relativistic Au$+$Au collisions at RHIC. Unfortunately, as we show below through the present Comment, there are many questions which are either ignored or not clarified properly in the works of~\cite{must,mustho}.

\section*{Notations}
Although the emphasis of~\cite{must} was on the massive charmed quark yet our discussion of various controversial questions becomes easiest in the massless case. Employing $c=1$ units let the leading parton moving in the transverse, i.e., x direction be associated with mass $m=0$, initial momentum $p_0$, initial energy $E_0=\mid p_0\mid$, travel time $t$, distance covered $L=t$, residual momentum variable $p=p_0-\delta p$, residual energy variable $E=\mid p \mid=E_0-\epsilon$, collisional momentum loss variable $\delta p$, energy loss variable $\epsilon$, drag coefficient $A(t)$, diffusion coefficient $D_F(t)$, useful integral $G(t)$, average momentum $\langle p\rangle_t$, useful kinematic symbol $\eta(p,t)$, and time dependent variance $W(t)$ defined by
$$
\begin{array}{lcl}
G(t)= exp\left\{-\int_0^t\,d\,t^{'}\, A(t^{'})\right\},\quad 
\eta=p-\langle p\rangle_t
\end{array}\eqno{(C1)}
$$
$$
\begin{array}{lcl}
W(t)=4\,G^2(t)\,\int_0^t\,d\,t^{'}\,D_F(t^{'})/G^2(t^{'}).
\end{array}\eqno{(C2)}
$$
{\it The first lacuna of~\cite{must,mustho} is that they contain mix-up of notations like $p+\epsilon$ at several spots [e.g., eq. (3) of ref. (1)] and also they fail to mention the fact that $E\neq \mid p\mid$, $\epsilon\neq \delta p$, $L\neq t$ in the massive ($m\neq 0$) parton case}. One may argue that, since jet quenching is a high-energy phenomenon with $p>5$ GeV typically, hence the charm quark mass $m\approx 1.5$ GeV can be ignored. This argument, however, is untenable because the fundamental quantity derived in~\cite{must} was the probability distribution $D_I(p,t)$ which must be computed even down to $p=0$ and plotted from $p=1$ GeV onwards [Fig. 4 of ref. 1]. In these regions the charm quark mass cannot be neglected, i.e., the care $E\neq|p|$ and $L\neq t$ should have been exercised in~\cite{must}. 

\section*{Fokker-Planck equation}

Refs.~\cite{must,mustho} are based on the fundamental transport equation [see eq. (35) of ref. 1]
$$
\begin{array}{lcl}
\frac{\partial D_I}{\partial t}=A\,\frac{\partial}{\partial p}(p\,D_I)+D_F\,\frac{\partial^2 D_I}{\partial p^2}\quad;\quad 0<t<\infty,
\end{array}\eqno{(C3)}
$$
where the transport coefficients are treated as momentum-independent according to the classical view and a suffix $I$ has been attached to the one-body distribution $D(p,t)$ for the sake of clarity. Imposing $\delta(p-p_0)$ initial condition the unique solution of (C3) is found to be [see eq. (42) of ref. 1]
$$
\begin{array}{lcl}
D_I(p,t)=\frac{1}{\sqrt{\pi\,W(t)}}\,exp\left\{-\frac{\eta^2}{W(t)}\right\}
\quad;\quad -\infty<p<\infty,
\end{array}\eqno{(C4)}
$$
where the exponential contains purely the second power of $\eta$. Unfortunately, the conventional version (C3) of the FP equation cannot properly describe relativistic brownian particle. The ($1+1$)-dimensional covariant Langevin dynamics~\cite{dunk} leads to a new FP equation in the massive case as 
$$
\begin{array}{lcl}
\frac{\partial D_H}{\partial t}=\frac{\partial}{\partial p}\{A\,p\,D_H+B\,E\,\frac{\partial D_H}{\partial p}\}
\end{array}\eqno{(C5)}
$$ 
where $D_H(p,t)$ is the H\"{a}nggi-D\"{u}nkel distribution function, $A(t)$ and $B(t)$ may be taken as momentum-independent coefficients, and the replacement $E\rightarrow|p|\rightarrow|\eta|$ is to be done in the massless limit. {\it The second lacuna of~\cite{must,mustho} is that they did not employ the relativistically  logical transport equation (C5) and also did not seek different type of time-dependent solution $D_H$ in which the exponential may contain $\mid \eta \mid$ linearly.}

\section*{Probability function in $p$ versus $E$}
 Even if the solution (C4) in momentum variable is assumed to be valid yet care must be exercised while transforming to the energy language. We know that both the points $\pm p$ are mapped into the same $E=\mid p\mid$. Hence, by standard mathematical statistics, the probability distribution in $E$ would read
$$
\begin{array}{lcl}
D_{II}(E,L)=D_I(p,L)+D_I(-p,L)\quad;\quad 0<E<\infty.
\end{array}\eqno{(C6)}
$$
which is automatically even in $p$ and preserves the total probability in the sense that  $\int_{-\infty}^{\infty}\,d\,p\,D_I(p,t)=\int_0^{\infty}\,d\,E\,D_{II}(E,L)=1.$ 

Unfortunately, for non-zero $\langle p\rangle_t$, the piece $D_I(p,t)=D_{III}(E,L)$ itself is not even in $p$, it cannot represent the probability function over the positive $E$ axis, and its normalization integral is awkward namely
$$
\begin{array}{lcl}
\int_0^{\infty}dp\,D_I(p,t)=\int_0^{\infty}dE\,D_{III}(E,L)=\frac{1}{2}\,erfc\left(\frac{-\langle p\rangle_t}{\sqrt{W}}\right)
\end{array}\eqno{(C7)}
$$
with erfc being the complimentary error function~\cite{abram}. {\it The third lacuna of~\cite{must,mustho} is that they wrongly assumed the energy-space probability function to be simply $D_{III}(E,L)=D_I(p,t)$ [see eq. (28) of ref. 2] which contradicts the logical expression (C6) and also yields awkward normalization (C7).}

\section*{Unphysical asymptotic form}

Suppose the philosophy of~\cite{must,mustho} is applied to a thought situation where the massless Brownian particle is embedded in a huge thermal bath at finite ambient temperature $T_{\infty}$. Keeping in mind the definitions (C2,C4) we expect that at instants large compared to relaxation time
$$
\begin{array}{lcl}
G(t)\rightarrow 0\quad;\quad \langle p\rangle_t\rightarrow 0\quad;\quad W(t)\rightarrow W_{\infty},
\end{array}\eqno{(C8)}
$$
$$
\begin{array}{lcl}
D_I(p,t)\rightarrow\frac{1}{\sqrt{\pi\,W_{\infty}}}\,exp\left\{\frac{-p^2}{W_{\infty}}\right\}
\neq \frac{1}{2\,T_{\infty}}\,exp\left\{\frac{-\mid p\mid}{T_{\infty}}\right\},
\end{array}\eqno{(C9)}
$$
{\it The fourth lacuna of ~\cite{must,mustho} is that the equilibrium distribution of their test particles does not tend towards the massless Maxwellian form as $t\rightarrow \infty$ even if Bose/Fermi statistics are ignored.} It is known that the correct asymptotic Maxwellian is retrieved if stationary solution is sought for the D\"{u}nkel-H\"{a}nggi~\cite{dunk} version of the FP equation (C5).  

\section*{Mean energy and loss}

Suppose the energy space distribution function for $0<E<\infty$ is just assumed to be $D_{III}(E,L)=D_I(p,t)$ (which it is NOT). Then, by making the transformation
$$
\begin{array}{lcl}
\sigma=(p-\langle p\rangle_t)/\sqrt{W}\quad;\quad \sigma_0=-\langle p\rangle_t/\sqrt{W},
\end{array}\eqno{(C10)}
$$
the average energy of the Brownian particle would read
$$
\begin{array}{lcl}
\langle E\rangle_{III}=\int_0^{\infty}\,d\,p\,p\,D_I(p,t)/\int_0^{\infty}\,d\,p\,D_I(p,t)=\\
\left\{\langle p\rangle_t\,erfc(\sigma_0)+\sqrt{W/\pi}\,exp(-\sigma_0^2)\right\}/erfc(\sigma_0),
\end{array}\eqno{(C11)}
$$
where erfc stands for the complementary error function~\cite{abram}. Clearly, $\langle E\rangle_{III}\approx \langle p\rangle_t\approx p_0\,G(t)$ only if $\mid \sigma_0\mid=\mid\langle p\rangle_t/\sqrt{W}\mid\gg 1$. {\em The fifth lacuna of~\cite{must,mustho} is that, instead of (C11), they report the average energy as simply $E_0\,G(t)$} [see eq. (31) of ref. 2] without mentioning any restriction on the ratio $\mid \langle p\rangle_t/\sqrt{W}\mid$. Of course, if the correct distribution $D_{II}(E,L)$ defined by (C6) were used the corresponding mean energy would become $\langle E\rangle_{II}=E_0\,G(t)$ rigorously.  

\section*{Convolution integral for hadron spectrum}

Let the shorthand notation $\Sigma(E,L)=dN/d^2p_T$ denote the $P_T$ distribution of hadrons produced by a jet of energy $E$ after covering a distance $L$. Then, assuming again the probability distribution to be $D_{III}(E,L)$ (which it is NOT) the medium modification over the vacuum spectrum is given by
$$
\begin{array}{lcl}
\Sigma^{med}(E,L)=\int_0^{\infty}\,d\epsilon\,\epsilon\,D_{III}(E,L,E_0=E+\epsilon)\,\Sigma^{vac}(E_0).
\end{array}\eqno{(C12)}
$$
which is a function of two variables $E$ and $L$ since $\epsilon$ has been integrated out.
Employing a linear Taylor expansion of $\Sigma^{vac}(E+\epsilon)$ around $E$ Mustafa-Thoma [see eq. (3) of ref. 1] arrived at the estimate
$$
\begin{array}{lcl}
\Sigma_{MT}^{med}\approx\Sigma^{vac}(E+\Delta E)\quad;\quad \Delta E =E_0\{1-G(L)\}.
\end{array}\eqno{(C13)}
$$

We wish to point out that the above estimate is unreliable because the experimentally parametrized light hadron production cross-section at RHIC energy namely~\cite{mustho}
$$
\begin{array}{lcl}
\Sigma^{vac}(E_0)\propto\left(1+\frac{E_0}{1.75\,{\rm GeV}}\right)^{-8},
\end{array}\eqno{(C14)}
$$
is a rapidly varying function of $E_0$. It is much safer to compute the integral (C12) by Simpson quadrature which reveals that the ratio $R\equiv\Sigma_{MT}^{med}/\Sigma_{Simpson}^{med}$, for $E=5\,{\rm GeV}$, is close to 1 at $L=1\,{\rm fm}$ but becomes about $170$ at $L=10\,{\rm fm}$. (see Fig. 1 for a more detailed view). {\it The sixth lacuna of~\cite{must,mustho} is that their Taylor expansion result (C13) grossly overestimates the convolution integral for large track lengths.}
\begin{figure}[tbp]
\begin{center}
{\includegraphics[scale=0.8]{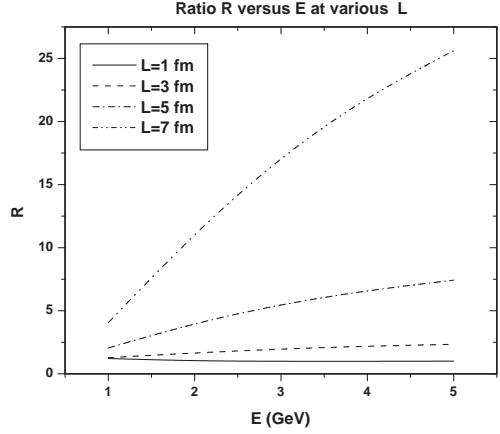}} 
\caption{The ratio $R\equiv\Sigma_{MT}^{med}/\Sigma_{Simp}^{med}$ plotted as a function of parton energy E at various track lengths $L$. The curve corresponding to $L=10$ fm has not been shown because $R$ shoots to $170$ for $E=5$ GeV.}
\end{center}
\end{figure} 

An important observation must be made at this juncture. The above discussion along with Fig. 1 is the easiest way to check the inadequacy of the MT approximation (C13) based on $\Delta E$, without getting bogged-down with experimental complications. If, however, actual RHIC data on jet quenching are to be fitted then due consideration must be given to the temperature $T\geq0.2$ GeV, plasma life time $t_{life}\leq 10$ fm/c, cylindrical geometry of Bjorken expansion, production configuration $(r,\phi)$ of the leading parton in the transverse plane, jet track lengths $L(r,\phi)\leq 10$ fm etc. Even then, the detailed convolution integral should be computed using Simpson's quadrature at every stage rather than the $\Delta E$ approximation. 
\section*{Conclusions}
In view of the recent RHIC experiments the subject of jet quenching has become a very important research area because one must know the precise energy loss suffered by the leading parton due to the collisional encounters and radiative bremsstrahlung. For calculating the collisional energy loss one should start from a FP equation consistent with special relativity and work out all properties of the distribution function $D(p,t)$ in conceptually valid manner. In this context refs.~\cite{must,mustho} give excellent physical picture of the mechanisms but, unfortunately, most of their algebraic equations as well as numerical graphs seem to be unreliable. We will address these issues in detail in a subsequent paper.

\section*{Acknowledgement}
M. Mishra and R. K. Dubey are grateful to the Council of Scientific and Industrial Research (CSIR), New Delhi for financial assistance.

\end{document}